\documentclass[10pt]{article}
\usepackage{amsmath,amsthm,amsfonts,amssymb,graphicx}
\usepackage[colorlinks=true,citecolor=blue,pdfpagemode=UseNone,pdfstartview=FitH]{hyperref}

\allowdisplaybreaks[4]

\theoremstyle{plain}

\theoremstyle{definition}

\theoremstyle{remark}

\title{Comment on Glenn Shafer's ``Testing by betting''}
\author{Vladimir Vovk\thanks%
  {Department of Computer Science,
  Royal Holloway, University of London,
  Egham, Surrey, UK.
  E-mail: \href{mailto:v.vovk@rhul.ac.uk}{v.vovk@rhul.ac.uk}.}}

\begin{document}
\maketitle

\begin{abstract}
  This note is my comment on Glenn Shafer's discussion paper ``Testing by betting'' \cite{Shafer:2020-local},
  together with two online appendices comparing p-values and betting scores.

    \medskip

    \noindent
    The  version of this note at \url{http://alrw.net/e} (Working Paper 8) is updated most often.
\end{abstract}

\section*{Main comment}
\addcontentsline{toc}{section}{Main comment}

Glenn Shafer's paper is a powerful appeal for a wider use of betting ideas and intuitions in statistics.
He admits that p-values will never be completely replaced by betting scores,
and I discuss it further in Appendix~A
(one of the online appendices, also including Appendix~G and \cite{Vovk:B}, that I have prepared to meet the word limit).
Both p-values and betting scores generalize Cournot's principle \cite{Shafer:2007-local},
but they do it in their different ways, and both ways are interesting and valuable.

Other authors have referred to betting scores as Bayes factors \cite{Shafer/etal:2011}
and e-values \cite{Vovk/Wang:arXiv1912a-local,Grunwald/etal:arXiv1906}.
For simple null hypotheses, betting scores and Bayes factors indeed essentially coincide
\cite[Section~1, interpretation 3]{Grunwald/etal:arXiv1906},
but for composite null hypotheses they are different notions,
and using ``Bayes factor'' to mean ``betting score'' is utterly confusing to Bayesians
\cite{Robert:2011-local}.
However, the Bayesian connection still allows us
to apply Jeffreys's (\cite{Jeffreys:1961}, Appendix B) rule of thumb to betting scores;
namely, a p-value of 5\% is roughly equivalent to a betting score of $10^{1/2}$,
and a p-value of 1\% to a betting score of 10.
This agrees beautifully with Shafer's rule (6), which gives, to two decimal places:
\begin{itemize}
\item
  for $p=5\%$, $3.47$ instead of Jeffreys's $3.16$ (slight overshoot);
\item
  for $p=1\%$, $9$ instead of Jeffreys's $10$ (slight undershoot).
\end{itemize}

The term ``e-values'' emphasizes the fundamental role of expectation in the definition of betting scores
(somewhat similar to the role of probability in the definition of p-values).
It appears that the natural habitat for ``betting scores'' is game-theoretic
while for ``e-values'' it is measure-theoretic \cite{Shafer:personal};
therefore, I will say ``e-values''
in Appendices~G and~A and in \cite{Vovk:B} (another appendix),
which are based on measure-theoretic probability.

In the online appendix \cite{Vovk:B} I give a new example
showing that betting scores are not just about communication;
they may allow us to solve real statistical and scientific problems
(more examples are given in the comment by my co-author Ruodu Wang \cite[463--464]{Shafer:2020-local}).
David Cox \cite{Cox:1975} discovered that splitting data at random
not only allows flexible testing of statistical hypotheses
but also achieves high efficiency.
A serious objection to the method is that different people analyzing the same data
may get very different answers
(thus violating ``inferential reproducibility'' \cite{Goodman/etal:2016,Held/Schwab:2020-local}).
Using e-values instead of p-values remedies the situation.

\subsection*{Acknowledgments}

Thanks to Ruodu Wang for useful discussions
and for sharing with me his much more extensive list of advantages of e-values.
This research has been partially supported by Amazon, Astra Zeneca, and Stena Line.

\appendix
\section*{Appendix G\quad Comparison with Good's rule of thumb}
\addcontentsline{toc}{section}{Appendix G\quad Comparison with Good's rule of thumb}
\renewcommand{\thesection}{G}

This online appendix to my main comment \cite[445--446]{Shafer:2020-local}
has been written after the other two online appendices,
and it is not referred to from the journal version.

Shafer's \cite[(6)]{Shafer:2020-local} preferred way of transforming p-values to e-values is
\[
  S(p)
  =
  \frac{1}{\sqrt{p}}
  -
  1
\]
(in Shafer's notation $S$ is a function of an observation $y$ rather than the p-value $p$,
but let me make it a function of $p$).
This agrees well not only with Jeffreys's but also with Good's \cite[Appendix~IV]{Good:1958} rule of thumb.
According to Good,
$S(p)$ should lie in the range
\[
  \left(
    \frac{1}{30p},
    \frac{3}{10p}
  \right)
\]
when $0.001<p<0.2$
(which Good felt were the values of $p$ that are usually of most practical interest).
For this range of $p$, Shafer's interval for $S(p)p$ is
\[
  \left\{
    \sqrt{p}-p:
    p \in (0.001,0.2)
  \right\}
  \approx
  (0.031,0.247),
\]
which is close to Good's interval
\[
  \left(
    \frac{1}{30},
    \frac{3}{10}
  \right)
  \approx
  (0.033,0.300).
\]

\section*{Appendix A\quad Cournot's principle, p-values, and e-values}
\addcontentsline{toc}{section}{Appendix A\quad Cournot's principle, p-values, and e-values}
\renewcommand{\thesection}{A}

This online appendix is based, to a large degree, on Glenn Shafer's ideas about the philosophy of statistics.
After a brief discussion of p-values and e-values as different extensions of Cournot's principle,
I list some of their advantages and disadvantages.

\subsection{Three ways of testing}

\begin{figure}[htbp]
\begin{center}
  \includegraphics[width=0.5\textwidth]{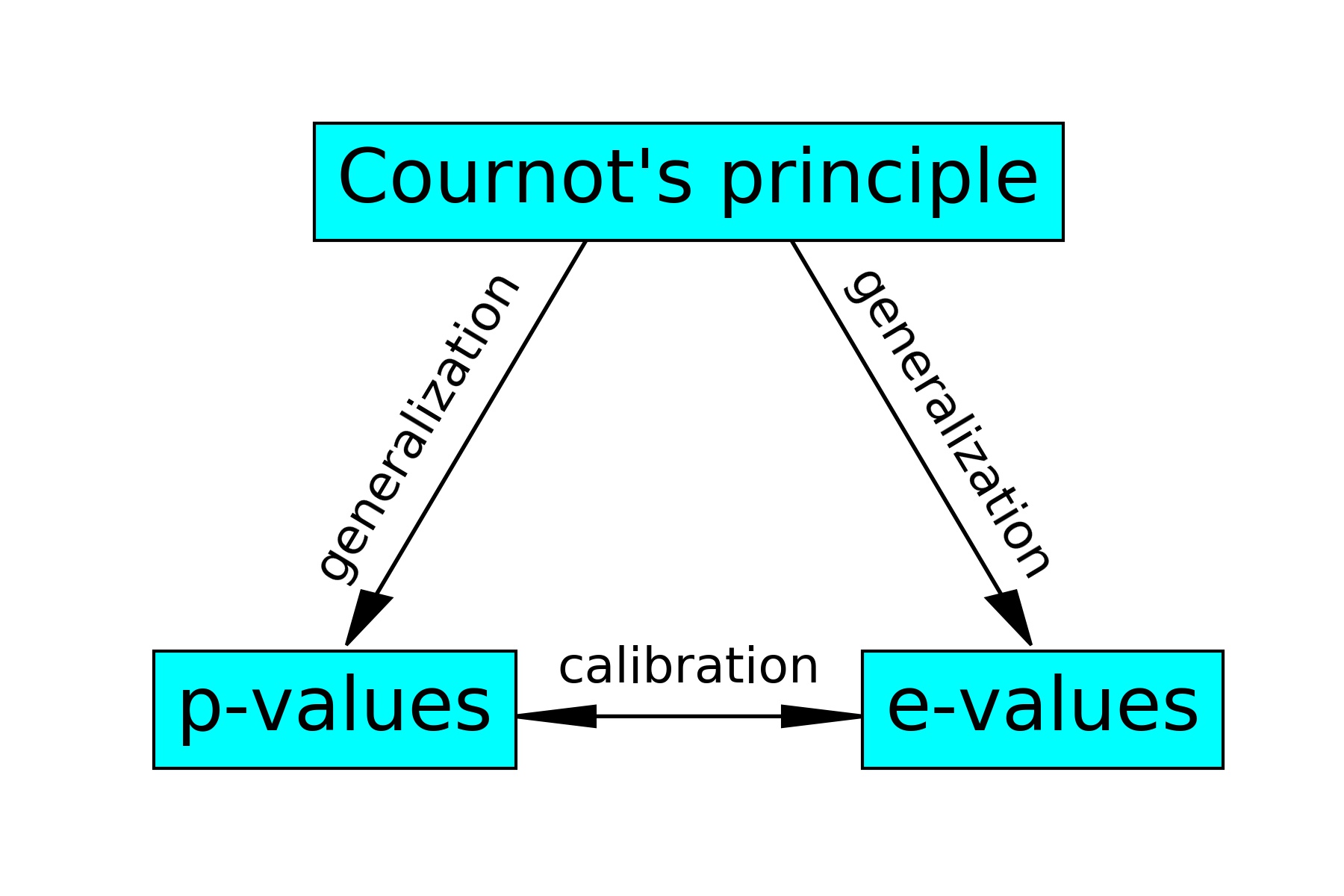}
\end{center} 
\caption{Cournot's principle and its two generalizations}\label{fig:Cournot}
\end{figure}

Both p-values and e-values are developments of Cournot's principle \cite{Shafer:2007-local},
which is referred to simply as the standard way of testing
in Shafer's \cite[Section 2.1]{Shafer:2020-local}.
If a given event has a small probability, we do not expect it to happen;
this is Cournot's bridge between probability theory and the world.
(This bridge was discussed already by James Bernoulli \cite{Bernoulli:1713};
Cournot's \cite{Cournot:1843} contribution was to say that this is the \emph{only} bridge.)
See Figure~\ref{fig:Cournot}.

Cournot's principle requires an \emph{a priori} choice of a rejection region $E$.
Its disadvantage is that it is binary:
either the null hypothesis is completely rejected or we find no evidence whatsoever against it.
A \emph{p-variable} is a nonnegative random variable $\mathsf{p}$ such that, for any $\alpha\in(0,1)$,
$P(\mathsf{p}\le\alpha)\le\alpha$;
one way to define p-variables is via Shafer's (3).
An \emph{e-variable} is a nonnegative random variable $\mathsf{e}$ such that $\mathbf{E}_P(\mathsf{e})\le1$;
one way to define e-variables is via Shafer's first displayed equation in Section 2.
In p-testing, we choose a p-variable $\mathsf{p}$ in advance and reject the null hypothesis $P$
when the observed value of $\mathsf{p}$ (the \emph{p-value}) is small,
and in e-testing, we choose an e-variable $\mathsf{e}$ in advance and reject the null hypothesis $P$
when the observed value of $\mathsf{e}$ (the \emph{e-value}) is large.
In both cases,
binary testing becomes graduated:
now we have a measure of the amount of evidence found against the null hypothesis.

We can embed Cournot's principle into both p-testing,
\[
  \mathsf{p}(y)
  :=
  \begin{cases}
    \alpha & \text{if $y\in E$}\\
    1 & \text{if not},
  \end{cases}
\]
and e-testing (as Shafer \cite[Section 2.1, (1)]{Shafer:2020-local} explains),
\[
  \mathsf{e}(y)
  :=
  \begin{cases}
    1/\alpha & \text{if $y\in E$}\\
    0 & \text{if not},
  \end{cases}
\]
where $\alpha:=P(E)$.

There are numerous ways to transform p-values to e-values (to \emph{calibrate} them)
and essentially one way ($e\mapsto1/e$) to transform e-values to p-values,
as discussed in detail in \cite{Vovk/Wang:arXiv1912b-local}.
The idea of calibrating p-values originated in Bayesian statistics
(\cite[Section 4.2]{Berger/Delampady:1987}, \cite[Section 9]{Vovk:1993logic}, \cite{Sellke/etal:2001}),
and there is a wide range of admissible calibrators.
Transforming e-values into p-values is referred to as \emph{e-to-p calibration} in \cite{Vovk/Wang:arXiv1912b-local},
where $e\mapsto1/e$ is shown to dominate any e-to-p calibrator \cite[Proposition 2.2]{Vovk/Wang:arXiv1912b-local}.

Moving between the p-domain and e-domain is, however, very inefficient.
Borrowing the idea of ``round-trip efficiency'' from energy storage,
let us start from the highly statistically significant ($\le1\%$) p-value $0.5\%$,
transform it to an e-value using Shafer's \cite[(6)]{Shafer:2020-local} calibrator
\[
  S(0.005)
  =
  \frac{1}{\sqrt{0.005}} - 1
  \approx
  13.14,
\]
and then transform it back to a p-value using the only admissible e-to-p calibrator:
$1/13.14\approx0.076$.
The resulting p-value of $7.6\%$ is not even statistically significant ($>5\%$).

\subsection{Some comparisons}

Both p-values and e-values have important advantages,
and I think they should complement (rather than compete with) each other.
Let me list a few advantages of each that come first to mind.
Advantages of p-values:
\begin{itemize}
\item
  P-values can be more robust to our assumptions (perhaps implicit).
  Suppose, for example, that our null hypothesis is simple.
  When we have a clear alternative hypothesis (always assumed simple) in mind,
  the likelihood ratio has a natural property of optimality as e-variable
  (Shafer \cite[Section 2.2]{Shafer:2020-local}),
  and the p-variable corresponding to the likelihood ratio as test statistic
  is also optimal (Neyman--Pearson lemma \cite[Section 3.2, Theorem 1]{Lehmann/Romano:2005}).
  For some natural classes of alternative hypotheses,
  the resulting p-value will not depend on the choice of the alternative hypothesis in the class
  (see, e.g., \cite[Chapter 3]{Lehmann/Romano:2005} for numerous examples;
  a simple example can be found in \cite[Section 4]{Vovk:B}).
  This is not true for e-values.
\item
  There are many known efficient ways of computing p-values for testing nonparametric hypotheses
  that are already widely used in science.
\item
  In many cases, we know the distribution of p-values under the null hypothesis: it is uniform on the interval $[0,1]$.
  If the null hypothesis is composite, we can test it by testing the simple hypothesis of uniformity for the p-values.
  A recent application of this idea is the use of conformal martingales
  for detecting deviations from the IID model \cite{Vovk:arXiv1906-full}.
\end{itemize}
Advantages of e-values (starting from advantages mentioned by Shafer \cite[Section 1]{Shafer:2020-local}):
\begin{itemize}
\item
  As Shafer \cite{Shafer:2020-local} convincingly argues,
  betting scores are more intuitive than p-values.
  Betting intuition has been acclaimed as the right approach to uncertainty
  even in popular culture \cite{Duke:2018}.
\item
  Betting can be opportunistic, in Shafer's words \cite[Sections 1 and 2.2]{Shafer:2020-local}.
  Outcomes of experiments performed sequentially by different research groups
  can be combined seamlessly into a nonnegative martingale
  \cite{ter/Grunwald:arXiv1905} (see also \cite[Section 1]{Grunwald/etal:arXiv1906}).
\item
  Mathematically, averaging e-values still produces a valid e-value,
  which is far from being true for p-values \cite{Vovk/Wang:2020}.
  This is useful in, e.g., multiple hypothesis testing \cite{Vovk/Wang:arXiv1912b-local}
  and statistical testing with data splitting \cite{Vovk:B}.
\item
  E-values appear naturally as a technical tool when applying the duality theorem
  in deriving admissible functions for combining p-values
  \cite{Vovk/etal:arXiv2007}.
\end{itemize}
\end{document}